\documentclass{aa}
\usepackage{amssymb}
\usepackage{epsf}
\def\astroncite#1#2{#2}
\def\aap{A\& A}
\def\pasp{Proc.Astr.Soc.Pacific}

\def\mnras{MNRAS}
\def\apj{ApJ}
\def\aj{AJ}

\def\erg{\hbox{erg}}
\def\sec{\hbox{s}}

\def\gram{\hbox{g}}
\def\cm{\hbox{cm}}

\def\AU{\hbox{AU}}

\def\therm{\hbox{\scriptsize abs}}
\def\irr{\hbox{\scriptsize irr}}
\def\emit{\hbox{\scriptsize emit}}
\def\dltime{{\tilde t}}
\def\dttime{{\bar t}}

\def\hshp{\chi}
\def\therm{\hbox{\scriptsize therm}}
\def\dyn{\hbox{\scriptsize dyn}}

\def\instab{\hbox{\scriptsize inst}}
\def\sigmar{\sigma_{\hbox{\tiny R}}}
\def\Hs{H_{\hbox{\scriptsize s}}}
\def\Hp{H_{\hbox{\scriptsize p}}}
\def\Te{T_{\hbox{\scriptsize e}}}
\def\massu{m_{\hbox{\scriptsize u}}}
\def\mug{\mu_{\hbox{\scriptsize g}}}
\def\heatcap{\Gamma}
\def\comma{\,,}
\def\fullstop{\,.}
\def\thttle{Are passive protostellar disks stable to self-shadowing?}
\begin{document}
\thesaurus{02.01.2,08.03.4,08.06.2,08.16.5,13.09.6}
\title{\thttle}
\author{C.P. Dullemond}
\authorrunning{Dullemond}
\titlerunning{\thttle} 
\institute{Max Planck Institut f\"ur Astrophysik, Karl
Schwarzschild Strasse 1,\\ D--85748 Garching, Germany; e--mail:
dullemon@mpa-garching.mpg.de}
\date{DRAFT, \today}

\maketitle

\begin{abstract}
The uniqueness and stability of irradiated flaring passive protostellar
disks is investigated in the context of a simplified set of equations for
the vertical height $H$ as a function of radius $R$. It is found that the
well-known flaring disk solution with $H\propto R^{9/7}$ is not unique.
Diverging solutions and asymptotically conical ($H\propto R$) solutions are
also found. Moreover, using time-dependent linear perturbation analysis, it
is found that the flaring disk solution may become unstable to
self-shadowing. A local enhancement in the vertical height alters the
functional form of irradiation grazing angle, and causes the 'sunny side' of
the enhancement to grow and the 'shadow side' to collapse in a run-away
fashion.
%
%
This instability operates in regions of the disk in which the cooling
time is much shorter than the vertical sound crossing time, which may occur
in the outer regions of the passive irradiated disk if dust and gas are
sufficiently strongly thermally coupled. Processes that may stabilize the
disk, which include active accretion, irradiation from above (e.g.~a
scattering corona) and low disk optical depth, are likely to operate only at
small or at large radius. The simple analysis of this Letter therefore
suggests that the instability may alter the flaring disk structure at
intermediate radii (between the actively accreting and fast rotating inner
regions and the optically thin outer regions).
%
\end{abstract}


\section{Introduction}
It has become widely accepted that the infrared excess observed in the
spectral energy distribution of many T-Tauri stars is caused by thermal
emission from a circumstellar dusty disk (Adams \& Shu,
\cite{adamsshu:1986}; Adams, Lada \& Shu,
\cite{adamsladashu:1987}). However, the spectra predicted by theoretical
models of both accretion-powered disks (Lynden-Bell \& Pringle,
\cite{lyndenpring:1974}) and flat reprocessing disks (Friedjung,
\cite{friedjung:1985}) have an infrared spectrum that goes as $\lambda
F_\lambda\propto \lambda^{-1.33}$. This is steeper than the $\lambda
F_\lambda\propto \lambda^{-0.5\;\hbox{\scriptsize to}\;-1.0}$ typically
observed from T-Tauri stars (Rucinski, \cite{rucinski:1985}; Rydgren \& Zak,
\cite{rydgrenzak:1987}). A natural way to explain the relatively warm dust
at large radii (which is needed to increase the slope of the model
spectrum), is to invoke a flaring disk geometry, in which $H/R\propto
R^{\gamma}$, with $\gamma>0$ (Kenyon \& Hartmann,
\cite{kenyonhart:1987}). Because of the angle between the disk surface and
the star, the disk intercepts considerable stellar flux even at large radii,
and therefore acquires a shallower temperature profile than the $T\propto
R^{-3/4}$ found from the accreting and/or flat disk models. The flaring
index $\gamma$ can be determined self-consistently by equating the
intercepted flux with the emitted blackbody flux, under the assumption that
the disk is vertically isothermal. These models yield $\gamma=2/7$ (Chiang
\& Goldreich, \cite{chianggold:1997}, henceforth CG97; D'Alessio et
al.~\cite{dalesscanthart:1999}, henceforth DCHLC99). Apart from producing
the flatter SEDs, it was recognized that the optically thin surface layers
of the disk will have much higher temperatures than the interior (Calvet et
al.\cite{calvetpatino:1991}; Malbet \& Bertout \cite{malbetbertout:1991};
CG97). In addition to producing an extra component in the SED, this
optically thin layer is also able to explain certain dust features in
emission which would otherwise be expected in absorption.  A more direct
piece of evidence for the flaring nature of protostellar disks was provided
by the HST images of HH30 (Burrows et al.~\cite{burrowsetal:1996}).

Despite the successes of the flaring disk model in explaining observations
of T-Tauri stars, there are still some theoretical issues that remain to be
addressed. In this paper we address two questions: are the flaring disk
models stable against perturbations in vertical height and temperature, and
are they unique solutions to which a time-dependent evolution of the passive
disk settles down? The issue of stability has been addressed before by
DCHLC99. They considered perturbations in the temperature of the disk, and
found that they damp out as they propagate inward. However, their analysis
is based on the assumption that the disk vertical height quickly responds to
changes in the temperature, or in other words that the vertical sound
crossing time $t_{\dyn}$ is much smaller than the cooling time
$t_{\therm}$. This is true for the inner regions of the disk, but breaks
down at large radii. In this Letter we study the other extreme case: the case
of $t_{\dyn}\gg t_{\therm}$, which may occur in the outer regions of
the disk. We start from the same equations as DCHLC99 and CG97, but replace
the static equation for the vertical disk height with a simplified one-zone
dynamic model. We derive the pertubation equations and dispersion relations
from them, and find that under these circumstances the disk becomes unstable.

We will start the analysis in Section \ref{sec-static} with a description of
the family of solutions of a static irradiated optically thick disk. The
flared disk model mentioned above is just one of these solutions. The other
solutions either diverge at, or are conical ($H\propto R$) beyond some
radius.  In Section \ref{sec-timedep} we will discuss the stability under
the two extreme conditions ($t_{\therm}\ll t_{\dyn}$ and $t_{\therm}\gg
t_{\dyn}$). In Section \ref{sec-conditions} we discuss the consequences of 
this instability, and when it is expected to show up.

\section{Static equations for an irradiated disk}\label{sec-static}
\begin{figure}
\epsfxsize=8.5cm\epsfysize=6.9cm\epsfbox{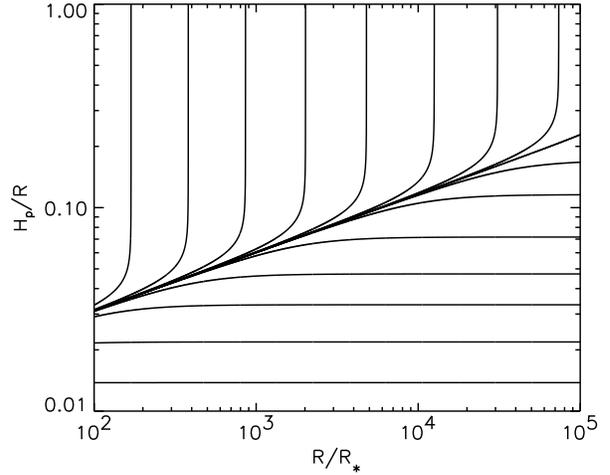}
\caption{The family of static irradiated passive disk solutions according
to Eq.(\ref{eq-dimless-ode-chianggold}) for $C=1\times 10^{14}$.}
\label{fig-all-sol}
\end{figure}

A disk with flaring angle $\alpha$ intercepts a flux of
$F_{\irr}=\alpha(R_{*}/R)^2\sigmar T_{*}^4\;\erg\cm^{-2}\sec^{-1}$. The
flaring angle $\alpha$ is defined as the angle between the grazing incident
stellar light and the surface of the flaring disk, and is assumed to be
small. If we assume the disk to be vertically isothermal, the emitted flux
is $F_{\emit}=\sigmar \Te^4$, where $\Te$ is the temperature of the
disk. Equating $F_{\irr}$ with $F_{\emit}$ yields
\begin{equation}
\Te = \alpha^{1/4}
\left(\frac{R_{*}}{R}\right)^{1/2} T_{*} \fullstop
\end{equation}
When $R\gg R_{*}$ the flaring angle is given by
\begin{equation}\label{eq-def-flare-angle}
\alpha = R\frac{d}{dR}\left(\frac{\Hs}{R}\right) \comma
\end{equation}
where $\Hs$ is the height of the surface of the disk above the midplane.
For a static solution the vertical pressure balance equation for an
isothermal disk yields a Gaussian density profile of the form
\begin{equation}\label{eq-rho-vert-profile}
\rho = \frac{\Sigma}{\Hp\sqrt{2\pi}}\exp\left(-\frac{z^2}{2\Hp^2}\right)
\comma
\end{equation}
where $\Sigma$ is the surface density, and $\Hp$ is the pressure scale height
\begin{equation}\label{eq-vert-hydrostat}
\Hp = \sqrt{\frac{k \Te R^3}{\mug \massu G M_{*}}}\comma
\end{equation}
where $\mug$ is the mean molecular weight and $\massu$ is the unit atomic
mass. The surface height $\Hs$ and the pressure scale height $\Hp$ are
related by a number of the order of a few, dependent on the opacity. We
assume this to be approximately a constant $\hshp\equiv \Hs/\Hp$. CG97 take
$\hshp\simeq 4$ for their analytic calculation of the disk structure.
If we eliminate $\Te$ from the above set of equations we arrive at
\begin{equation}\label{eq-ode-chianggold}
\frac{1}{\hshp}\left(\frac{\mug \massu G M_{*}}{k T_{*} R_{*}}\right)^4
\left(\frac{R_{*}}{R}\right)^2 \; \left(\frac{\Hp}{R}\right)^8
=R\frac{d}{dR}\left(\frac{\Hp}{R}\right)
\fullstop
\end{equation}
Defining the dimensionless quantities $r\equiv R/R_{*}$ and 
$h\equiv \Hp/R$, one arrives at
\begin{equation}\label{eq-dimless-ode-chianggold}
r\frac{dh(r)}{dr} = \frac{C}{r^2} h(r)^8
\comma
\end{equation}
where $C$ is 
\begin{equation}\label{eq-def-c-const}
C=\frac{1}{\hshp}\left(\frac{\mug \massu G M_{*}}{k T_{*} R_{*}}\right)^4
\fullstop
\end{equation}
This ordinary differential equation (ODE) has a 1-parameter family of 
solutions:
%
%
\begin{equation}
h(r)=\left(\frac{7C}{2}\frac{1}{r^2}+K\right)^{-1/7}
\comma
\end{equation}
where $K$ is an arbitrary real constant. For $K<0$ the solutions diverge
at $r=\sqrt{-7C/2K}$, for $K>0$ the solutions become constant beyond
$r=\sqrt{7C/2K}$, and for $K=0$ one obtains a powerlaw solution.
These three classes of solutions are shown in Fig.(\ref{fig-all-sol}). 

The power law solution is given by
\begin{equation}\label{eq-sol-powerlaw}
h(r) = \left(\frac{2}{7C}\right)^{1/7}\,r^{2/7}\,\comma
\end{equation}
which is the solution given by CG97. All the other solutions asymptotically
tend to this power law solution, Eq.(\ref{eq-sol-powerlaw}), as $r\downarrow
0$. But as $r$ goes to infinity the solutions deviate from this power law at
some radius, and either diverge to $h\!\rightarrow\!\infty$ at a particular
radius, or asymptotically approach a conical geometry
$h(r)\!\rightarrow\!\hbox{const}$, i.e. $\Hs\propto R$. 
%

The direction in which the solutions deviate from each other is interesting.
Since the inner regions of the disk (small radii) cast their shadow over the
outer regions (large radii), the ``causality'' of the system is pointing
outwards. A change in the structure of the disk at small radii has its
repercussions on the disk at large radii, but not vice versa. In principle
this means that the (numerical) integration of the ODE
(Eq.\ref{eq-ode-chianggold}) must be done from inside-to-outside (i.e.  in
causal direction). However, in this direction the solutions tend to strongly
deviate from each other. A slight change in integration constant at the
start of the integration causes a very large difference at large radii.  For
stable integration of the ODE (Eq.\ref{eq-ode-chianggold}) one should place
the boundary condition at the outside and integrate inwards. 
Though mathematically correct, it is physically meaningless, and may hint 
to an intrinsic instability of the time-dependent equations. 
\section{Time-dependent equations for irradiated disk}\label{sec-timedep}
The stability of the flaring (power law) solution can be studied analytically
using time-dependent linear perturbation analysis. We study two regimes, one
of which has been studied before by DCHLC99. 

\subsection{Temperature perturbations at hydrostatic equilibrium}
In the analysis of DCHLC99 the cooling time scale of the
disk was assumed to be much larger than the orbital time scale, which is
true for the inner regions of the disk. We may then assume that the disk
will always be in hydrostatic equilibrium, and we follow perturbations of
the temperature. Following DCHLC99 we write
\begin{equation}
\heatcap\frac{d \Te}{dt} = F_{\irr} - F_{\emit}
\comma
\end{equation}
where $\heatcap$ is the thermal heat capacity. After substituting
Eqs.(\ref{eq-def-flare-angle},\ref{eq-vert-hydrostat},\ref{eq-def-c-const}),
the definitions of $F_{\irr}$ and $F_{\emit}$, and defining the perturbation
$\Te(r,t)=\Te^0(r)(1+\tilde\psi(r,t) r^2)$, where $\Te^0(r)$ is the
temperature belonging to the static flaring disk solution
Eq.(\ref{eq-sol-powerlaw}), one finds the following perturbation
equation (see DCHLC99 for details):
\begin{equation}\label{eq-temppert}
\frac{\partial \tilde\psi}{\partial\dttime} = 
\frac{7}{4}\left(\frac{\Te^0}{T_{*}}\right)^3r\frac{\partial\tilde\psi}{\partial r}
\comma
\end{equation}
where $\dttime\equiv (\sigmar T_{*}^3/\heatcap)t$. Eq.(\ref{eq-temppert}) is
an advection equation. Apparently a perturbation moves inwards, and its
amplitude damps, since $\tilde \psi$ is conserved in amplitude along the
inward moving characteristic, and therefore the comoving amplitude of
$\tilde\psi r^2$ becomes smaller (DCHLC99).

\subsection{Hydrodynamic perturbations at thermal equilibrium}
When the cooling time of the disk is much shorter than the orbital time, one
can assume that the disk is always in thermal equilibrium, and the disk
height will react to changes in the irradiation flux in a hydrodynamic way,
instead of being in vertical hydrostatic balance. In principle one should
solve the time-dependent vertical hydrodynamics of the disk (1-D
hydrodynamics). But for the present analysis we simplify the picture, and
{\em assume} that the density profile is always given by
Eq.(\ref{eq-rho-vert-profile}), but with $\Hp$ a function of time. This
assumption can be justified, since in the early (linear) stages of a
possible instability one has $dv_z/dt\propto z$, which preserves the
Gaussian shape of the density profile. By neglecting quadratic terms in the
vertical velocity, one can then derive
\begin{equation}
\frac{d^2 \Hp}{dt^2} = \frac{k \Te}{\mug \massu}\frac{1}{\Hp} - 
\frac{G M_{*}}{R^3} \Hp
\fullstop
\end{equation}
When we define the dimensionless time $\dltime\equiv \sqrt{G
M_{*}/R_{*}^3}t$, a dimensionless form of the above equation is
\begin{equation}\label{eq-timdep-dimless}
\frac{\partial^2 h(r,\dltime)}{\partial\dltime^2} = 
\frac{C^{-1/4}}{r^{5/2} h(r,\dltime)}
\left(r\frac{\partial h(r,\dltime)}{\partial r}\right)^{1/4}
\!\!- \frac{h(r,\dltime)}{r^3}
\comma
\end{equation}
with $C$ as defined in Eq.(\ref{eq-def-c-const})

We now analyze the time-dependent behavior of linear perturbations
of the power law solution (Eq.\ref{eq-sol-powerlaw}). We take
\begin{equation}
h(r,\dltime) =\left(\frac{2}{7C}\right)^{1/7}\!\!r^{2/7}\,[1+\psi(r,\dltime)]
\fullstop
\end{equation}
Substituting this in Eq.(\ref{eq-timdep-dimless}) yields (exactly):
\begin{equation}
\frac{\partial^2\psi}{\partial \dltime^2} =\frac{1}{r^3}\left\{
\frac{1}{(1+\psi)}\left(1+\psi+\frac{7}{2}r\frac{\partial\psi}{\partial
r}\right)^{1/4}\!\!\! - (1+\psi)\right\}
\fullstop
\end{equation}
For $\psi\ll 1$ this becomes
\begin{equation}
\frac{\partial^2\psi}{\partial \dltime^2} = 
- \frac{7}{8r^3}\left\{2\psi-r\frac{\partial\psi}{\partial r} \right\}
\fullstop
\end{equation}
This equation has solutions of the form
\begin{equation}
\psi = A\, r^2 e^{\sigma\dltime} \sin\left(\omega\dltime-kr^3+\phi_0\right)
\comma
\end{equation}
with the following dispersion relations:
\begin{eqnarray}
\omega^2 &=& \sigma^2 \comma\\
k &=& -(16/21)\;\omega\sigma \fullstop
\end{eqnarray}
%
For inwards propagating waves (i.e.~$k/\omega<0$) one has $\sigma>0$, and
hence an exponentially growing mode\footnote{After submission 
the author became aware of work by E.~Chiang (to be published) which reports on
similar results.}. The instability growth rate is inversely proportional
to the wavelength of the perturbation, which indicates that the instability
will be dominated by the shortest wavelengths. It should be noted that the
validity of the equations breaks down at the scale of the disk vertical
height. For modes with larger $k$ one should take radial radiative diffusion
into account. This is, however, beyond the scope of the present set of
equations.

\section{Conditions for instability}\label{sec-conditions}
The instability only sets in under certain physical conditions. The most
obvious conditions for this instability to occur are:
\begin{enumerate}
\item\label{item-dyntherm} Radiative cooling/heating time scale short 
compared to vertical sound crossing time, i.e. $t_{\dyn}\gg t_{\therm}$.
\item Thermal dust-gas coupling is strong enough to keep the {\em gas} cooling
time shorter than the vertical sound crossing time.
\item The equatorial temperature is dominated by irradiation, i.e.~viscous
dissipation by accretion is small in comparison.
\item The disk is optically thick to starlight in radial (grazing) direction.
\item The irradiation occurs through flaring; i.e. the reflected stellar
flux from a scattering corona is weak in comparison to the direct
interception of stellar light through flaring. 
\end{enumerate}
We have not investigated whether disk models that obey all the above points,
but are vertically optically thin in the near/mid infrared, are also
unstable. 

Condition \ref{item-dyntherm} is fulfilled for $R\gg R_{\instab}$,
where $R_{\instab}$ can be derived by equating the vertical sound crossing
time with the thermal time: $t_{\dyn}=t_{\therm}$. We have
\begin{eqnarray}
t_{\dyn} &\simeq& \sqrt{\frac{R^3}{GM_{*}}}\comma \\
t_{\therm} &\simeq& \frac{k\Sigma}{2\massu\sigmar \Te^3}\fullstop
\end{eqnarray}
The surface density $\Sigma(R)$ (in units of $\gram/\cm^{2}$) is a free
function of the model, as long as the disk remains optically thick to the
incident stellar radiation. Equating $t_{\dyn}=t_{\therm}$ yields the radius
$R_{\instab}$ beyond which the instability may set in:
\begin{equation}
\frac{R_{\instab}}{\AU} = 3\times 10^{-5}\;
\left(\frac{M_{*}}{M_{\odot}}\right)^{13/3}
\left(\frac{L_{*}}{L_{\odot}}\right)^{-4}
\left(\frac{\Hs}{\Hp}\right)^{-4}
\Sigma^{14/3}
\comma
\end{equation}
where $L_{*}$ is the luminosity of the star. 
%

The instability found here is an instability of
the highly simplified analytic flaring disk models. Including more realistic
physics could perhaps stabilize the disk. Some elements of realistic physics
that should be considered in future work before one can claim that the
instability is real are:
\begin{enumerate}
\item 2-D axisymmetric hydrodynamics.
\item Radial radiative transfer of stellar radiation instead of
the inclination angle formula.
\item 1-D vertical radiative transfer, because
the top layers have a higher temperature (see CG97).
\item Radial diffusion of radiation, or even fully 2-D 
radiative transfer, because radial exchange of energy
might counter-act the shadowing effect which causes the instability.
\end{enumerate}


\section{Discussion}
This Letter presents an analysis of the structure and stability of
irradiated non-accreting protostellar disks based on a highly simplified set of
equations. In the context of these equations it is concluded that the
flaring non-accreting reprocessing disk model is unstable when the vertical
sound crossing time is larger than the heating/cooling time scale. It should
be noted, however, that detailed multidimensional numerical modeling is
required to verify whether this instability is real or merely an artifact of
the over-simplified equations.
%
%

The instability has a simple physical interpretation. Consider a flaring
disk solution, and perturb it with an enlargement of the scale height at some
point. The `sunny side' of the hill will get overheated and expand
vertically in a run-away fashion. The `shadow side' will receive
insufficient radiation and collapse. As the perturbations grow into the
non-linear regime they start to cast a `real' shadow ($\alpha<0$) over the
disk at larger radii, thus completely depriving this part of the disk of
irradiation. At this point the validity of the simplified equations breaks
down, and what happens in this non-linear regime is unclear. Disk self
irradiation (e.g.~Bell \cite{bell:1999}) might become important at this
stage. 

The reason why we find an instability, while DCHLC99 found only damping
modes is because they considered only perturbations in the disk where the
thermal time scale exceeds the vertical sound crossing time scale.  While
for actively accreting disks this condition is always satisfied, for passive
disks the reverse may be true. This was recognized by DCHLC99, but an
analysis of modes on a dynamic time scale in the outer regions of the
protostellar disks was not carried out, hence their different conclusion.

\begin{acknowledgements}
I wish to thank H.~Spruit, C.~Dominik, P.~Armitage, S.~Doty, J.~Papaloizou,
E.~v.~Dishoeck, G-J.~v.~Zadelhoff and the referee P.~D'Alessio for useful
suggestions and comments.
\end{acknowledgements}


\end{document}